\documentclass[12pt]{iopart}

\usepackage{iopams}
\usepackage{graphicx}
\usepackage{subfigure}
\usepackage{mcite}

\begin{document}

\title{Additional electron pairing in a \emph{d}-wave superconductor driven by nematic order}

\author{Jing-Rong Wang$^{1,2}$ and Guo-Zhu Liu$^{1,3}$ \footnote[0]{$^3$ Author to whom
any correspondence should be addressed.}}

\address{$^1$Department of Modern Physics, University of Science and
Technology of China, Hefei, Anhui, 230026, P.R. China}
\address{$^2$Max Planck Institut f$\ddot{u}$r Physik komplexer
Systeme, D-01187 Dresden, Germany} \ead{gzliu@ustc.edu.cn}

\begin{abstract}
We perform a non-perturbative analysis of the strong interaction
between gapless nodal fermions and the nematic order parameter in
two-dimensional $d_{x^2-y^2}$ superconductors. We predict that the
critical nematic fluctuation can generate a dynamical nodal gap if
the fermion flavor $N$ is smaller than a threshold $N_c$. Such gap
generation leads to an additional is-wave Cooper pairing
instability, which induces a fully gapped $d_{x^2-y^2}+is$
superconducting dome in the vicinity of the nematic quantum critical
point. The opening of a dynamical gap has important consequences,
including the saturation of fermion velocity renormalization, a weak
confinement of fermions and the suppression of observable
quantities.
\end{abstract}


\maketitle


\section{Introduction}

One of the most prominent properties of high-$T_c$ copper-oxide
superconductors is that they exhibit a number of long-range orders
upon changing the chemical doping, such as antiferromagnetism,
superconductivity, stripe, nematic state, and so on. The competition
and coexistence between superconductivity and other long-range
orders are believed to be fundamental issues in the studies of
high-$T_c$ superconductors. Among the orders that are in competition
with superconductivity, the nematic order, which spontaneously
breaks $C_4$ symmetry down to $C_2$ symmetry, has attracted special
theoretical and experimental interest in the past decade
\cite{Kivelson98, Kivelson03, FradkinA, FradkinB, Vojta09}.

In recent years, strong anisotropy in the electronic properties has
been observed in various experiments performed on
YBa$_{2}$Cu$_{3}$O$_{6+\delta}$ \cite{Ando02, Hinkov08, Daou10} and
Bi$_{2}$Sr$_{2}$CaCu$_{2}$O$_{8+\delta}$ \cite{Lawler10}. Such
strong anisotropy is universally attributed to the formation of an
electronic nematic state \cite{FradkinA,FradkinB} in these two
typical high-$T_c$ superconductors. It is very interesting to notice
that similar nematic states have also been observed in a list of
other correlated electron systems, including iron-based
superconductor \cite{Chuang10}, heavy fermion superconductor
\cite{Okazaki11}, Sr$_{3}$Ru$_{2}$O$_{7}$ superconductor
\cite{Borizi07}, and even semiconductor heterostructure
\cite{Cooper02}.

Motivated by the observed strong electronic anisotropy in high-T$_c$
superconductors, many researchers anticipate the existence of a
nematic quantum phase transition in these systems \cite{Kim08,
Huh08, Xu08, Fritz09, Wang11, Liu12, WangLiu}. Such nematic
transition and the associated nematic critical behaviors have been
investigated extensively in the past several years \cite{Kim08,
Huh08, Xu08, Fritz09, Wang11, Liu12, WangLiu}. It is well-known that
high-T$_c$ superconductors have a $d_{x^2 - y^2}$ energy gap, which
vanishes linearly at four nodal points $(\pm \frac{\pi}{2a},\pm
\frac{\pi}{2a})$. Due to this special property, gapless nodal
quasiparticles (qps) exist in the superconducting state even in the
low-energy regime. If a nematic phase transition occurs in the
superconducting dome, the fluctuation of nematic order parameter
couples to the gapless nodal qps. This coupling becomes singular at
the nematic quantum critical point (QCP), and can lead to unusual
behaviors \cite{Kim08, Huh08, Xu08, Fritz09, Wang11, Liu12,
WangLiu}.

Vojta \emph{et al.} first analyzed the effective field theory of the
coupling between nematic order and nodal qps by means of $\epsilon =
3-d$ expansion and found runaway behavior \cite{Vojta00A,Vojta00B}.
Later, perturbative expansion in powers of $1/N$ with $N$ being the
fermion flavor has been extensively applied to address this issue
\cite{Kim08, Huh08, Xu08, Fritz09, Wang11, Liu12, WangLiu}. For
instance, Kim \emph{et. al.} revealed a second-order nematic phase
transition after performing a large-$N$ analysis \cite{Kim08}. More
recent renormalization group calculations of Huh and Sachdev found a
novel fixed point that exhibits extreme fermion velocity anisotropy
\cite{Huh08}. Subsequent studies showed that such extreme anisotropy
can lead to a variety of nontrivial properties, such as non-Fermi
liquid behavior \cite{Xu08}, enhancement of thermal conductivity
\cite{Fritz09}, and suppression of superconductivity \cite{Liu12}.
The influence of weak quenched disorders on the nematic QCP was also
addressed \cite{Wang11}.

We should note that all the pervious field-theoretic analysis are
based on the conventional perturbative expansions
\cite{Vojta00A,Vojta00B, Kim08, Huh08, Xu08, Fritz09, Wang11,
Liu12}. The non-perturbative effects have not been seriously
addressed. To illustrate this issue, we now consider the $d$-wave
superconducting state, which has low-lying elementary excitations
with spectrum \cite{Durst} $E_{k} = \sqrt{(\epsilon_{k} - \mu)^{2} +
\Delta_{\mathbf{k}}^{2}}$, where the electron dispersion
$\epsilon_{k} = 2 t_{f}\left(\cos k_{x}a+\cos k_{y}a\right)$ and the
$d$-wave gap $\Delta_{\mathbf{k}} = \frac{1}{2}\Delta_{0}\left(\cos
k_{x}a - \cos k_{y}a\right)$. In the vicinity of the gap node
$(\frac{\pi}{2a}, \frac{\pi}{2a})$, one can linearize the spectrum
and obtain $E_{k} =
\sqrt{v_{F}^{2}k_{1}^{2}+v_{\Delta}^{2}k_{2}^{2}}$, where $k_{1} =
\left(k_{x}+k_{y}-\pi/a\right)/\sqrt{2}$, $k_{2} =
\left(k_{x}-k_{y}\right)/\sqrt{2}$. The fermion velocity of nodal
qps is defined as ${\mathbf v}_{\Delta} = \partial
\Delta_{k}/\partial \mathbf{k}$ and gap velocity ${\mathbf v}_F =
\partial \epsilon_{k}/\partial \mathbf{k}$. For the other three
nodes, the linearization can be performed analogously. In this
formalism, one starts from the following free fermion propagator,
\begin{eqnarray}
G_{0}(\omega,\mathbf{k}) = \frac{1}{-i\omega + v_F k_{1} \tau^z +
v_{\Delta} k_{2} \tau^x},\label{eqn:G0}
\end{eqnarray}
where $\tau^{x,z}$ are two standard Pauli matrices, and then
perturbatively calculates the fermion self-energy
$\Sigma(\omega,\mathbf{k})$ induced by the interaction with nematic
fluctuation. Generically, $\Sigma(\omega,\mathbf{k})$ can be
expanded as
\begin{eqnarray}
\Sigma(\omega,\mathbf{k}) = -i\Sigma_0\omega + \Sigma_1 v_F k_1
\tau^z + \Sigma_2 v_{\Delta} k_2 \tau^x,\label{eqn:SFEnergy}
\end{eqnarray}
where the functions $\Sigma_{0}$ and $\Sigma_{1,2}$ are the temporal
and spatial components respectively. The fermion damping effect is
encoded in $\Sigma_{0}$ \cite{Kim08}, whereas the velocity
renormalization can be obtained from $\Sigma_{0,1,2}$ \cite{Huh08}.
However, in principle there could be the fourth term, $m\tau^y$,
which is defined by the third Pauli matrix $\tau^y$ and corresponds
to a nonzero mass gap term of the nodal qps. This mass term can
never be obtained to any finite order of the perturbative expansion
of fermion self-energy, but may be dynamically generated if one
performs non-perturbative calculations.

Another motivation of studying the non-perturbative effects is to
examine the validity of the $1/N$ expansion. When performing the
standard perturbative expansion in powers of $1/N$, the flavor $N$
is usually supposed to be quite large \cite{Kim08, Huh08, Xu08,
Fritz09, Wang11, Liu12}. However, in the present nematic problem,
the physical flavor of nodal qps is $N = 2$, determined by the spin
degeneracy. It is very interesting, and even necessary, to go beyond
the perturbative $1/N$ expansion and testify whether the
non-perturbative effects give rise to any nontrivial phenomena those
cannot be captured by the usual perturbative calculations.

In this paper, we study dynamical gap generation of originally
gapless nodal qps due to nematic fluctuation by means of
non-perturbative expansion. With the help of Dyson-Schwinger (DS)
equation that connects the free and complete propagators of nodal
qps, we obtain a nonlinear gap equation of fermion mass $m$ in the
vicinity of nematic QCP. After solving this equation, we find that a
nonzero mass gap, $m\tau^y$, is dynamically generated when the
fermion flavor $N$ is below certain critical value $N_c$, i.e. $N <
N_c$. We demonstrate that the dynamical gap $m$ induced by nematic
order corresponds to a secondary $is$-wave Cooper pairing formation,
so the critical nematic fluctuation drives a transition from a pure
$d_{x^2 - y^2}$ superconducting state to a $d_{x^2 - y^2} + is$
superconducting state in the vicinity of nematic critical point. As
a consequence, the superconducting state is fully gapped and the
massive nodal qps are weakly confined by a logarithmic potential.
Moreover, the dynamical gap leads to saturation of fermion velocity
renormalization and strong suppression of some observable
quantities.

The rest of the paper is organized as follows. In Section 2, we
perform a non-perturbative analysis by means of DS equation method
and examine whether a dynamical gap can be generated by the critical
nematic fluctuation. In Section 3, we discuss the physical
implications of dynamical gap generation. In Section 4, we briefly
summarize our results and comment on two interesting issues
concerning the validity of $1/N$ expansion and disorder effects.

\section{Non-perturbative calculations and gap generation}

The effective low-energy model describing the coupling between
nematic order and gapless nodal qps has already been derived and
extensively studied in previous publications \cite{Kim08, Huh08,
Xu08, Fritz09, Wang11, Liu12}. This effective model is composed of
the following three parts \cite{Kim08, Huh08, Xu08, Fritz09, Wang11,
Liu12}
\begin{eqnarray}
S &=& S_{\psi} + S_{\phi} + S_{\psi\phi}.\label{eqn:ActionTotal}
\end{eqnarray}

The free action for the nodal qps is
\begin{eqnarray}
S_{\psi} &=&
\int\frac{d^{2}\mathbf{k}}{(2\pi)^{2}}\frac{d\omega}{2\pi}
\Psi^{\dagger}_{1\sigma}(-i\omega+v_{F}k_{1}\tau^{z} +
v_{\Delta}k_{2}\tau^{x})\Psi_{1\sigma} \nonumber \\
&& + \int\frac{d^{2}\mathbf{k}}{(2\pi)^2}\frac{d\omega}{2\pi}
\Psi^{\dagger}_{2\sigma}(-i\omega+v_{F}k_{2}\tau^{z} +
v_{\Delta}k_{1}\tau^{x})\Psi_{2\sigma}.\label{eqn:ActionFermion}
\end{eqnarray}
Here, the nodal qps are described by Nambu spinors $\Psi_{1,2}$,
defined as
$\Psi_{1\sigma}=(f_{1\sigma},\epsilon_{\sigma,-\sigma}f_{3-\sigma}^{\dag})^{T}$
and
$\Psi_{2\sigma}=(f_{2\sigma},\epsilon_{\sigma,-\sigma}f_{4-\sigma}^{\dag})^{T}$,
where $\epsilon_{\sigma,-\sigma} = -\epsilon_{-\sigma,\sigma}$ with
spin indices $\sigma,-\sigma$. The four fermion operators $f_{1}$,
$f_{2}$, $f_{3}$, and $f_{4}$ represent gapless nodal qps excited
from four nodal points $(\frac{\pi}{2a},\frac{\pi}{2a})$,
$(-\frac{\pi}{2a},\frac{\pi}{2a})$,
$(-\frac{\pi}{2a},-\frac{\pi}{2a})$, and
$(\frac{\pi}{2a},-\frac{\pi}{2a})$, respectively. The fermion flavor
is determined by the spin degeneracy, so apparently $N = 2$.

The action for the nematic order parameter $\phi$ takes the standard
form,
\begin{eqnarray}
S_{\phi} = \int
d^2\mathbf{x}d\tau\left[\frac{1}{2}(\partial_{\tau}\phi)^2 +
\frac{c^2}{2}(\nabla\phi)^2 +
\frac{r}{2}\phi^2+\frac{u}{4!}\phi^4\right],\label{eqn:ActionBoson}
\end{eqnarray}
where $\phi$ is a real scalar field since the nematic transition is
accompanied by a discrete symmetry breaking (i.e., Ising-type). The
interaction between gapless nodal qps and nematic order parameter is
described by a Yukawa coupling term \cite{Vojta00A,Vojta00B}
\begin{eqnarray}
S_{\psi\phi} = \int d^2\mathbf{x}d\tau\{\lambda
\phi(\Psi^{\dagger}_{1\sigma}\tau^{x}\Psi_{1\sigma} +
\Psi^{\dagger}_{2\sigma}\tau^{x}\Psi_{2\sigma})\},\label{eqn:ActionCoupling}
\end{eqnarray}
where $\lambda$ is the coupling constant.

According to the standard perturbation theory, one would make
perturbative expansion in powers of coupling constant $\lambda$.
However, as revealed by the renormalization group analysis of
Ref.~\cite{Vojta00A,Vojta00B}, $\lambda$ tends to diverge in the
low-energy region and there is no stable fixed point. It turns out
that $\lambda$ is not an appropriate expanding parameter in the
present interacting system. It was later realized that a reasonable
route to access such model is to fix the parameter $\lambda$ at
certain finite value \cite{Kim08, Huh08, Sachdevbook}, and then to
perform perturbative expansion in powers of $1/N$.

\begin{figure}[htbp]
\center
\includegraphics[width=5.3in]{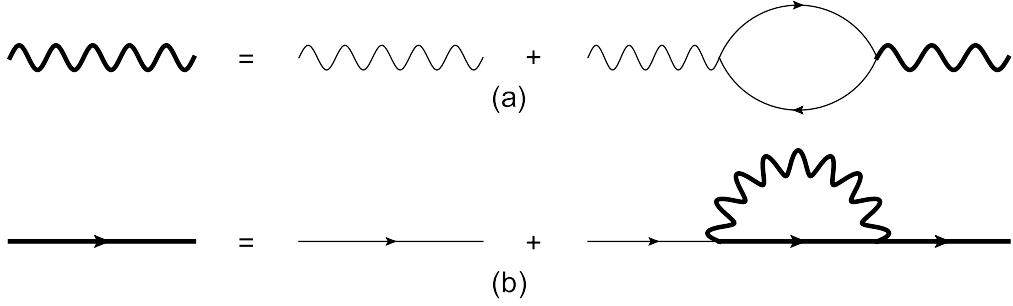}
\caption{(a) Dynamical screening of the propagator of nematic order
parameter; (b) DS equation of the fermion propagator. For
simplicity, the vertex corrections are not included.}
\label{Fig:feynman}
\end{figure}

In order to carry out analytical calculations, it is convenient to
assume a general fermion flavor $N$. The free propagator of nodal
qps is
\begin{equation}
G_{\Psi_{1\sigma}}^0(\omega,\mathbf{k}) =
\frac{1}{-i\omega+v_{F}k_{1}\tau^{z} +
v_{\Delta}k_{2}\tau^{x}}\label{eqn:G01}
\end{equation}
for $\Psi_{1\sigma}$, and
\begin{equation}
G_{\Psi_{2\sigma}}^0(\omega,\mathbf{k}) =
\frac{1}{-i\omega+v_{F}k_{2}\tau^{z} +
v_{\Delta}k_{1}\tau^{x}}\label{eqn:G02}
\end{equation}
for $\Psi_{2\sigma}$, respectively. The free propagator of the
nematic order parameter is
\begin{eqnarray}
D_0(\Omega,\mathbf{q}) = \frac{1}{\Omega^2 + \mathbf{q}^2 +
r}.\label{eqn:D0}
\end{eqnarray}
Due to the coupling between nematic fluctuation and nodal qps, the
nematic propagator can be dynamically screened, as shown in figure
\ref{Fig:feynman}(a), and becomes
\begin{eqnarray}
D(\Omega,\mathbf{q}) &=& \frac{1}{D_0^{-1}(\Omega,\mathbf{q}) +
\Pi(\Omega,\mathbf{q})} = \frac{1}{\Omega^2 + \mathbf{q}^2 + r +
\Pi(\Omega,\mathbf{q})},
\end{eqnarray}
where $\Pi(\Omega,\mathbf{q})$ is the polarization function. To the
leading order of the $1/N$ expansion, the polarization function is
given by
\begin{eqnarray}
\fl \Pi(\Omega,\mathbf{q}) &=& \lambda^2 \sum_{\sigma=1}^{N}
\sum_{i=1,2}
\int\frac{d^{2}\mathbf{k}}{(2\pi)^{2}}\frac{d\omega}{2\pi}
\mathrm{Tr}\left[\tau^{x}G^{0}_{\Psi_{i\sigma}}(\omega,\mathbf{k})
\tau^{x} G^{0}_{\Psi_{i\sigma}}(\omega+\Omega,
\mathbf{k+q})\right].\label{eqn:PolarizationA}
\end{eqnarray}
After straightforward calculations, it is easy to get
\begin{eqnarray}
\Pi(\Omega,\mathbf{q}) = \frac{N\lambda^2}{16v_F v_\Delta}
\frac{\Omega^2+v_{F}^{2}q_{1}^{2}}{Q_{1}} + (q_{1} \leftrightarrow
q_{2}),\label{eqn:PolarizationB}
\end{eqnarray}
where $Q_{1} = \sqrt{\Omega^2 + v_{F}^{2}q_{1}^{2} +
v_{\Delta}^{2}q_{2}^{2}}$. This polarization is linear in $|q|$, and
dominates over the kinetic term $q^2$ at low energies. We can drop
the $q^2$ term and write the effective nematic propagator as
\begin{eqnarray}
D(\Omega,\mathbf{q}) = \frac{1}{r +
\Pi(\Omega,\mathbf{q})}.\label{eqn:DressedD}
\end{eqnarray}

The free fermion propagator also receives corrections due to its
coupling with the nematic fluctuation. After including these
corrections, the complete propagator of $\Psi_{1\sigma}$ takes the
following general form
\begin{eqnarray}
G_{\Psi_{1\sigma}}(\omega,\mathbf{k}) &=& \frac{1}{-i\omega
A_{0}+v_{F}k_{1}A_{1}\tau^{z}+v_{\Delta}k_{2}A_{2}\tau^{x} +
m\tau^{y}},\label{eqn:Gfull}
\end{eqnarray}
where $A_{0,1,2}$ are wave-function renormalizations and $m \equiv
m(\omega,k_{1},k_{2})$ is a fermion gap. The mass gap term can never
be generated so long as the fermion self-energy is calculated
perturbatively using the free fermion propagator
$G^0(\omega,\mathbf{k})$. To examine the possibility of dynamical
gap generation, we should go beyond the perturbative level and
instead utilize the following self-consistent DS equation
\begin{eqnarray}
G_{\Psi_{1\sigma}}^{-1}(\varepsilon,\mathbf{p}) =
\left[G_{\Psi_{1\sigma}}^{0}(\varepsilon,\mathbf{p})\right]^{-1} -
\Sigma_{1\sigma}(\varepsilon,\mathbf{p})\label{eqn:Dyson}
\end{eqnarray}
where the self-energy is computed as follows,
\begin{eqnarray}
\Sigma_{1\sigma}(\varepsilon,\mathbf{p}) &=&
\lambda^2\int\frac{d\omega}{2\pi}
\frac{d^2\mathbf{k}}{(2\pi)^2}\tau^{x}
G_{\Psi_{1\sigma}}(\omega,\mathbf{k})\tau^{x}
\frac{1}{r+\Pi(\varepsilon - \omega,\mathbf{p} -
\mathbf{k})}.\label{eqn:SFEnergyConsistent}
\end{eqnarray}

Notice the non-perturbative feature of this approach is reflected in
the fact that the complete fermion propagator is used in the
right-hand side of equation (\ref{eqn:SFEnergyConsistent}). After
substituting equation (\ref{eqn:Gfull}) into the DS equation, one
can derive four self-consistently coupled equations of $A_{0,1,2}$
and $m$. Generically, the equations of $A_{0,1,2}$ can be expanded
in the form: $A_{0,1,2} = 1 + \mathcal{O}(1/N)$. To the leading
order of $1/N$ expansion, we assume that $A_{0} = A_{1} = A_{2} = 1$
and ignore all higher order corrections. To the leading order, the
gap equation is given by
\begin{eqnarray}
\fl m(\varepsilon,p_{1},p_{2}) &=&\lambda^2\int\frac{d\omega}{2\pi}
\frac{d^2\mathbf{k}}{(2\pi)^2}
\frac{m(\omega,k_{1},k_{2})}{\omega^2+v_{F}^{2}k_{1}^2 +
v_{\Delta}^{2}k_{2}^{2} + m^{2}(\omega,k_{1},k_{2})}
\frac{1}{r+\Pi(\epsilon-\omega,\mathbf{p}-\mathbf{k})}.\label{eqn:GapEqA}
\end{eqnarray}
If this equation has only vanishing solution, $m = 0$, then the
nematic fluctuation can not open any gap. A fermion gap is
dynamically generated once this equation develops a nontrivial
solution, i.e., $m \neq 0$. In contrast, if the free fermion
propagator $G_{\Psi_{1\sigma}}^{0}(\varepsilon,\mathbf{p})$ is
substituted into equation (\ref{eqn:SFEnergyConsistent}), one would
obtain the usual perturbative results of the fermion self-energy. In
such case, no dynamical fermion gap can be generated even after
including higher order corrections, i.e., $m \equiv 0$.

\begin{figure}[htbp]
\center
\includegraphics[width=4in]{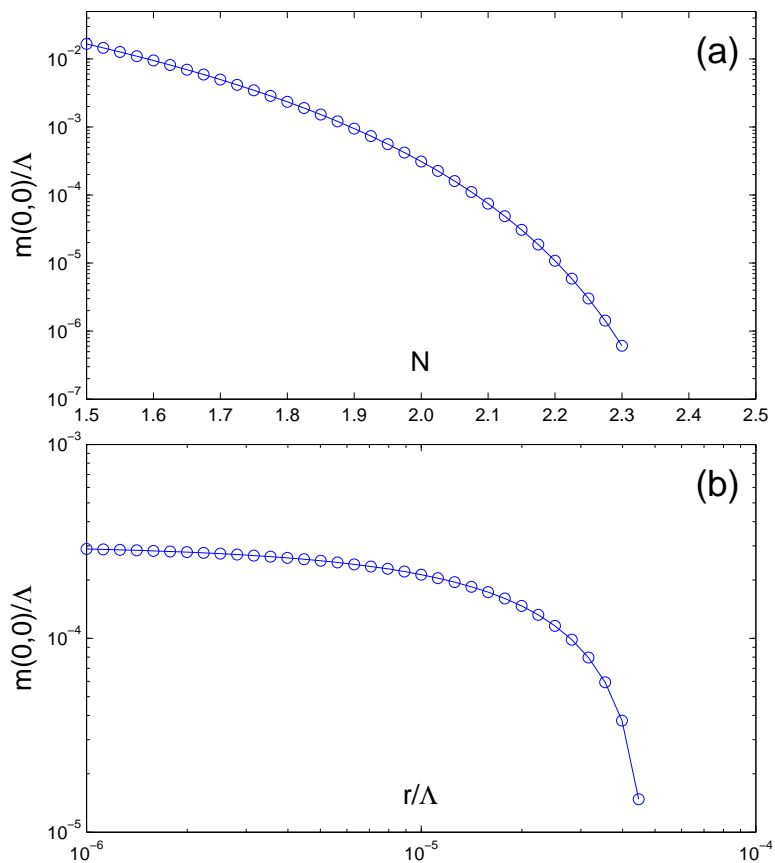}
\caption{(a) Relationship between dynamical gap $m(0,0)$ and fermion
flavor $N$ at nematic QCP, $r=0$; (b) Dependence of $m(0,0)$ on
tuning parameter $r$ for $N=2$. Dynamical gap is destroyed once $r$
exceeds certain critical value $r_c$.} \label{Fig:Ga0Nr}
\end{figure}

We now attempt to solve the complicated nonlinear equation
(\ref{eqn:GapEqA}). Due to the anisotropic nature of nematic
fluctuation, the integrations over $\omega$, $k_1$, and $k_2$ have
to be performed separately, which greatly increases the difficulty
of numerical computations. For simplicity, here we consider the
isotropic limit, $v_{F} = v_{\Delta}$. In this case, the dynamical
gap becomes $m(\varepsilon,|\mathbf{p}|)$, and the polarization is
simplified to
\begin{eqnarray}
\Pi(\Omega,\mathbf{q}) &=&
\frac{N\lambda^2}{16v_{F}^{2}}\frac{2\Omega^2 +
v_{F}^{2}|\mathbf{q}|^{2}}{\sqrt{\Omega^2+v_{F}^{2}|\mathbf{q}|^{2}}}.
\label{eqn:PolarizationC}
\end{eqnarray}

We first consider the nematic QCP and take $r = 0$. In this special
case, the pre-factor $\lambda^2$ on the right-hand side of equation
(\ref{eqn:GapEqA}) cancels exactly the factor $\lambda^2$ appearing
in the polarization function,
$\Pi(\epsilon-\omega,\mathbf{p}-\mathbf{k})$, so the gap equation
becomes
\begin{eqnarray}
\fl m(\varepsilon,|\mathbf{p}|) &=&
\frac{1}{N}\int\frac{d\omega}{2\pi} \frac{d^2\mathbf{k}}{(2\pi)^2}
\frac{m(\omega,|\mathbf{k}|)}{\omega^2+v_{F}^{2}|\mathbf{k}|^2 +
m^{2}(\omega,|\mathbf{k}|)} \frac{16 v_{F}^{2}
\sqrt{(\epsilon-\omega)^2 +
v_{F}^{2}|\mathbf{p}-\mathbf{k}|^{2}}}{2(\epsilon-\omega)^2 +
v_{F}^{2}|\mathbf{p}-\mathbf{k}|^{2}}.
\end{eqnarray}
This gap equation is independent of $\lambda$, and the critical
point of dynamical gap generation is therefore solely determined by
the flavor $N$. After numerical computations, we find a critical
fermion flavor $N_{c} \approx 2.4$ at $r=0$. The nodal qps remain
gapless, $m=0$, when $N
> N_c$, but acquire a nonzero dynamical gap, $m \neq 0$ when $N < N_c$.
figure \ref{Fig:Ga0Nr}(a) presents the dependence of the static gap
$m(\omega = 0,\mathbf{k} = 0)$ on flavor $N$. It is clear that the
dynamical gap decreases very rapidly as flavor $N$ grows, and
vanishes continuously as $N \rightarrow N_c$.

It is also interesting to examine how the tuning parameter $r$ of
nematic transition affects the dynamical gap generation. Actually,
if we move away from the nematic QCP, $r$ becomes finite and the
nematic fluctuation is no longer critical. For $r \neq 0$, the
coupling parameter $\lambda$ cannot be exactly canceled, but one can
absorb it into $r$ by taking $r \rightarrow r/\lambda^2$. We find
that the dynamical gap is significantly suppressed by growing $r$,
and completely destroyed once $r$ exceeds certain critical value
$r_c$, which is shown in figure \ref{Fig:Ga0Nr}(b). We therefore
conclude that the dynamical gap generation is mediated by the
critical fluctuation of nematic order parameter, and exists only in
the vicinity of the nematic QCP.

\section{Physical implications of dynamical gap}

The dynamically generated gap for gapless nodal qps can result in a
series of nontrivial physical consequences. In this section, we will
discuss the physical implications of the dynamical gap.

Once a nonzero dynamical gap $m$ is generated for the originally
gapless nodal qps, there will be an extra term that should be added
to the Hamiltonian:
\begin{eqnarray}
\fl H_{\mathrm{m}} &=& \int\frac{d^2\mathbf{k}}{(2\pi)^{2}}\{m
(\Psi^{\dagger}_{1\sigma}\tau^{y}\Psi_{1\sigma} +
\Psi^{\dagger}_{2\sigma}\tau^{y}\Psi_{2\sigma})\} \nonumber \\
\fl&=& \int\frac{d^2\mathbf{k}}{(2\pi)^{2}}\left\{im\left[\left(
f_{3\downarrow}^{\dag}f_{1\uparrow}^{\dag} +
f_{3\downarrow}f_{1\uparrow}
\right)+\left(f_{1\downarrow}^{\dag}f_{3\uparrow}^{\dag} +
f_{1\downarrow}f_{3\uparrow}\right) + \left(1 \leftrightarrow 2,3
\leftrightarrow 4\right)\right]\right\}.
\end{eqnarray}
One can immediately identify that such dynamically generated term
corresponds to the formation of singlet Cooper pairs between the
gapless nodal qps excited from opposite nodes. We therefore have
obtained a secondary, nematic-order driven, $is$-wave
superconducting instability on top of the pure $d_{x^2 - y^2}$
superconductivity.

\begin{figure}[ht]
\centering
   \includegraphics[width=4.1in]{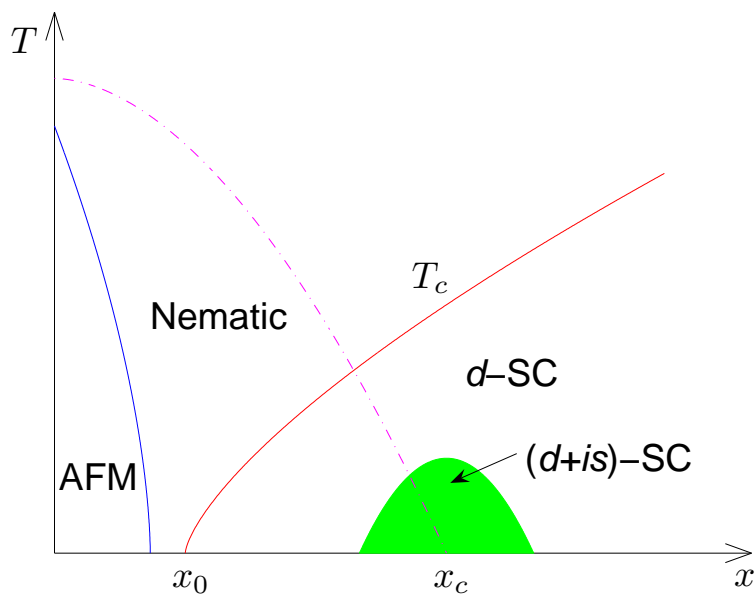}
\caption{Schematic phase diagram. $x_c$ denotes the nematic QCP. The
shadowed region around $x_c$ represents the emergent fully-gapped
$d_{x^2 - y^2} + is$ superconducting dome within a much larger, pure
$d_{x^2 - y^2}$ superconducting dome.}\label{Fig:phase}
\end{figure}

As already pointed out, this dynamical gap is opened only when $r$
is zero or very small, and is rapidly destroyed as $r$ increases,
which implies that the secondary $is$-wave superconductivity is
achieved only in the close vicinity of nematic QCP. Upon approaching
the nematic QCP, there is a zero-temperature phase transition from a
pure $d_{x^2 - y^2}$ superconducting state to a fully gapped $d_{x^2
- y^2} + is$ superconducting state. At finite temperature, $T \neq
0$, the critical nematic fluctuation is weakened strongly due to the
thermal screening effects, hence the dynamical nodal gap cannot
survive at high temperatures. According to these analysis, we now
can infer that a small $d_{x^2 - y^2} + is$ superconducting dome
emerges around the nematic QCP, which is schematically shown in
figure \ref{Fig:phase}. It is interesting to notice that such
nematic fluctuation-driven superconducting dome is analogous to that
is formed on the border of an antiferromagnetic quantum critical
point in the contexts of some heavy fermion compounds
\cite{heavyfermion, Stockert}. We also notice that the
non-perturbative effects of coupling between nodal qps and
fluctuating order parameter has been investigated in a physically
different context \cite{Khve}.

We next discuss the effects of nonzero dynamical gap on a number of
quantities. First of all, we consider the classical potential
between the nodal qps. For simplicity, let us assume a constant gap
$m$, which yields a new polarization
\begin{eqnarray}
\fl \Pi(\mathbf{q},\Omega) &=& \frac{N\lambda^2}{2\pi
v_{F}v_{\Delta}} \frac{\Omega^2+v_{F}^{2}q_{1}^{2}}{Q_{1}}
\left[\frac{1}{2}\frac{m}{Q_{1}} +
\left(\frac{1}{4}-\frac{m^2}{Q_{1}^{2}}\right)\arcsin\left(\frac{Q_{1}}
{\sqrt{4m^2 + Q_{1}^{2}}}\right)\right]\nonumber
\\
\fl &&+(q_{1}\leftrightarrow q_{2}).\label{eqn:PolarizationD}
\end{eqnarray}
In the low energy limit, it takes the simplified form
\begin{eqnarray}
\Pi(\mathbf{q},\Omega) \approx \frac{N\lambda^2}{12\pi v_F v_\Delta
m}(2\Omega^2+v_{F}^{2}|\mathbf{q}|^{2}).\label{eqn:PolarizationE}
\end{eqnarray}
Using this simplified polarization, it is easy to obtain an
effective potential
\begin{eqnarray}
V(\mathbf{R}) &\propto& \int\frac{d^2\mathbf{q}}{(2\pi)^{2}}
\frac{e^{i\mathbf{q}\cdot\mathbf{R}}}{\Pi(\mathbf{q})} \nonumber
\\
&\propto& \frac{6v_{\Delta}}{N\lambda^2
v_{F}}m\ln(mR)\label{eqn:Potential}
\end{eqnarray}
between two massive fermions \cite{Maris}. This potential grows
logarithmically as the distance $R$ is increasing, so the gapped
fermions are weakly confined. Such gap-induced fermion confinement
is similar to that in a physically analogous theory of QED$_{3}$
\cite{Maris}.

\begin{figure}[ht]
\centering
   \includegraphics[width=3.3in]{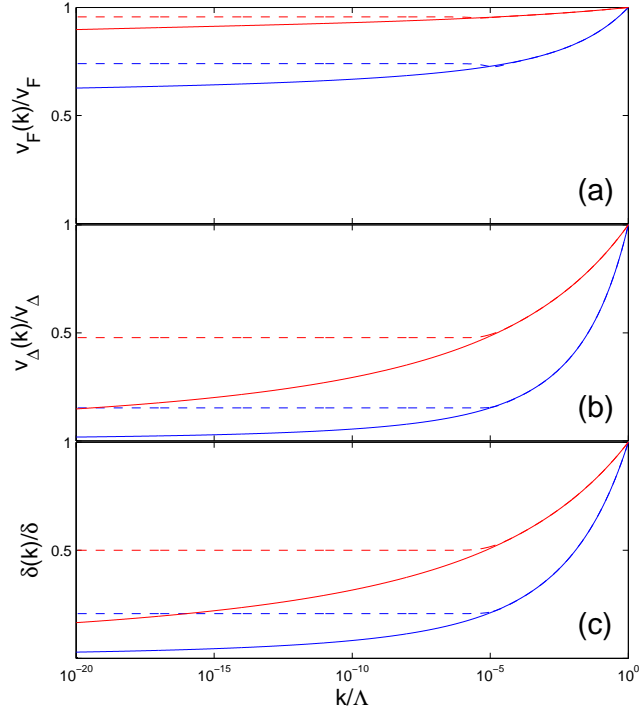}
\caption{Momentum dependence of fermion velocities $v_{F,\Delta}(k)$
and velocity ratio $\delta(k) = v_{\Delta}(k)/v_F(k)$. Here,
$v_{F,\Delta}$ and $\delta$ are bare values of the corresponding
quantities. Blue solid line: $\delta = 1$ and $m = 0$; Red Solid
line: $\delta = 0.1$ and $m = 0$; Blue dashed line: $\delta = 1$ and
$m/v_{F}\Lambda = 10^{-6}$; Red dashed line: $\delta = 0.1$ and
$m/v_{F}\Lambda = 10^{-6}$. }\label{Fig:VRG}
\end{figure}

When the nodal qps are massless, their velocities are renormalized
by the nematic fluctuation and thus become strongly
momentum-dependent,
\begin{eqnarray}
v_{F,\Delta} \rightarrow v_{F,\Delta}(k).\label{eqn:RunV}
\end{eqnarray}
The expressions of the renormalized velocities $v_{F,\Delta}(k)$ are
quite complicated and therefore not shown here, but can be easily
found in Refs.~\cite{Huh08, Wang11}. Both $v_{F}(k)$ and
$v_{\Delta}(k)$ vanish as $k \rightarrow 0$. However,
$v_{\Delta}(k)$ vanishes much more rapidly than $v_{F}(k)$, leading
to the so-called extreme velocity anisotropy \cite{Huh08, Xu08,
Fritz09, Wang11, Liu12},
\begin{eqnarray}
\delta(k) = \frac{v_{\Delta}(k)}{v_{F}(k)} \rightarrow
0.\label{eqn:RunDelta}
\end{eqnarray}
Nevertheless, once a finite fermion gap $m$ is opened, there is no
longer strong renormalization of velocities. As shown in figure
\ref{Fig:VRG}, the velocities $v_{F,\Delta}$ are still $k$-dependent
and decrease with decreasing $k$ at high energies, but saturate to
finite values once $k$ is smaller than the scale set by $m$.
Therefore, the singular velocity renormalization and the extreme
anisotropy are both prevented by fermion gap $m$.

The dynamical fermion gap has important impacts on a number of
observable quantities. For example, the density of state of nodal
qps is linear in energy, $\rho(\omega) \propto |\omega|$, when $m =
0$, but becomes $\rho(\omega) \propto |\omega| \theta\left(|\omega|
- m\right)$ when $m \neq 0$, which vanishes for $|\omega| < m$.
Accordingly, the specific heat of nodal qps is strongly suppressed
as $C_{V}(T) \propto m^4\exp(-m/T)/T^2$ in the low-temperature
region of $T \ll m$. Furthermore, the low temperature dc thermal
conductivity at finite $m$ is known to have the form \cite{Gusynin},
$\frac{\kappa}{T} = \frac{k_{B}^{2}}{3}(\frac{v_{F}}{v_{\Delta}}
+\frac{v_{\Delta}}{v_{F}}) \frac{\Gamma^{2}}{\Gamma^2+m^2}$, where
$\Gamma$ is the impurity scattering rate. In the massless limit, $m
= 0$, the thermal conductivity is a constant, $\frac{\kappa}{T} =
\frac{k_{B}^{2}}{3}(\frac{v_{F}}{v_{\Delta}} +
\frac{v_{\Delta}}{v_{F}})$, which is finite and impurity independent
\cite{Durst}. In contrast, once a fermion gap is generated, the
thermal conductivity is suppressed by the finite $m$.

Finally, notice that the nematic state is indeed equivalent to a
superconducting state with $d_{x^2 - y^2} + s$ gap, which was
pointed out in Ref.~\cite{Vojta00A,Vojta00B}. Therefore, in the
vicinity of a quantum critical point between a pure $d_{x^2 - y^2}$
superconducting state and a $d_{x^2 - y^2} + s$ superconducting
state, the singular fluctuation of $s$-wave order parameter can also
lead to a fully gapped $d_{x^2 - y^2} + is$ superconducting state,
provided that the flavor $N$ of nodal qps is smaller than the
corresponding critical value $N_c$.

\section{Summary and discussions}

In summary, we perform non-perturbative analysis within an effective
field theory of the strong interaction between the critical nematic
fluctuation and nodal qps in the context of \emph{d}-wave HTSCs. We
propose that a dynamical gap may be generated for the originally
gapless nodal qps in the vicinity of the nematic QCP. Such gap
generation is driven by the critical fluctuation of nematic order
parameter and corresponds to an additional $is$-wave Cooper pairing
instability. In the vicinity of the nematic QCP, there will be an
small emergent $d_{x^2 - y^2} + is$ superconducting dome. We also
discuss the physical implications of the dynamically generated gap,
and show that such gap leads to weak confinement of nodal qps,
saturation of velocity renormalization, and strong suppression of
several observable quantities.

According to our results, it turns out that the fermion flavor $N$
is a crucial parameter which determines the low-energy behaviors
caused by the nematic order. A critical value $N_c$ is found to
exist. When $N > N_c$, the non-perturbative effects of nematic
fluctuation is unimportant, so one can trust the results obtained by
perturbative calculations, such as extreme anisotropy \cite{Huh08}
and other unusual properties \cite{Kim08, Xu08, Fritz09, Wang11,
WangLiu, Liu12}. If $N < N_c$, however, the non-perturbative effects
become significant, and can drive an additional $is$-wave
superconducting pairing between the originally gapless nodal qps.

Our leading-order calculations found that $N_c \approx 2.4$, which
is larger than the physical flavor $N = 2$. It would be interesting
to study how $N_c$ is quantitatively affected by high order
corrections. In principle, it is straightforward to address this
issue by coupling the equations of wave function renormalizations
$A_{0,1,2}$ and vertex corrections to the gap equation.
Unfortunately, solving these coupled equations is a highly
challenging task because the integrations over three components of
momentum, $\omega, k_1, k_2$, have to be performed separately due to
the non-relativistic and spatially anisotropic feature of the
present system. It is quite difficult to get reliable numerical
solutions. We expect that large scale Monte Carlo simulations would
be utilized to investigate this problem and help to determine the
precise value of $N_c$.

Irrespective of whether our $N_c$ is precise or not, a general trend
can be deduced from our results: the conventional perturbative $1/N$
expansion should be reliable for large $N$, but it may fail to
capture some fundamental features of strongly interacting model for
small $N$ and non-perturbative analysis should be utilized instead.
In addition, our prediction of a nematic order-induced $d_{x^2 -
y^2} + is$ superconducting dome is novel and would shed light on the
investigation of nematic order in correlated electron systems.

In our present analysis, we have considered only clean $d$-wave
superconductors and ignored the disorder effects. The influence of
various quenched disorders on the stability of nematic QCP was
investigated in a recent paper \cite{Wang11}. As shown in this paper
\cite{Wang11}, the strong coupling between critical nematic
fluctuation and gapless nodal qps is actually not affected by weak
random gauge potential and weak random mass \cite{Wang11}. On the
contrary, random chemical potential is able to destroy nematic QCP
and thus can fundamentally change the whole picture. However, both
these conclusions and the analytical methods used in
Ref.~\cite{Wang11} are valid only in the particular case that the
non-perturbative effects of nematic fluctuation are unimportant and
all the nodal qps are strictly gapless. Once the non-perturbative
effect becomes strong enough to generate a dynamical fermion gap,
the influence of disorders might be quite different. Generically,
the dynamical gap generation and disorder scattering can affect each
other \cite{LiuWang}, so they should be investigated
self-consistently, as we have done in a physically similar context
\cite{LiuWang}. Nevertheless, this issue is beyond the scope of the
present paper, and would be addressed in the future. In any case, we
believe the results presented in this paper are reliable in clean
\emph{d}-wave superconductors and pointed out an interesting new
possibility regarding the exotic effects of nematic order.

\ack{We thank Jing Wang for helpful discussions. J.R.W. acknowledges
support by the MPG-CAS doctoral promotion programme. G.Z.L.
acknowledges financial support by the National Natural Science
Foundation of China under grants No. 11074234 and No. 11274286.}


\section*{References}

\begin{thebibliography}{99}

\bibitem{Kivelson98}
Kivelson S A, Fradkin E and Emery V J 1998 \emph{Nature} (London)
{\bf 393} 550

\bibitem{Kivelson03}
Kivelson S A, Bindloss I P, Fradkin E, Oganesyan V, Tranquada J M,
Kapitulnik A and Howald C 2003 \emph{Rev. Mod. Phys.} {\bf 75} 1201

\bibitem{FradkinA}
Fradkin E, Kivelson S A, Lawler M J, Eisenstein J P and Mackenzie A
P 2010 \emph{Annu. Rev. Condens. Matter Phys.} {\bf 1} 153

\bibitem{FradkinB}
Fradkin E 2012 in \emph{Modern Theories of Many-Particle Systems in
Condensed Matter Physics} edited by Cabra D C, Honecker A and Pujol
P Lecture Notes in Physics Vol. 843 (Berlin Heidelberg:
Springer-Verlag)

\bibitem{Vojta09}
Vojta M 2009 \emph{Adv. Phys.} {\bf 58} 699

\bibitem{Ando02}
Ando Y, Segawa K, Komiya S and Lavrov A N 2002 \emph{Phys. Rev.
Lett.} {\bf 88} 137005

\bibitem{Hinkov08}
Hinkov V, Haug D, Fauqu\'{e} B, Bourges P, Sidis Y, Ivanov A,
Bernhard C, Lin C T and Keimer B 2008 \emph{Science} {\bf 319} 597

\bibitem{Daou10}
Daou R, Chang J, LeBoeuf D, Cyr-Choini\`{e}re O, Lalibert\'{e} F,
Doiron-Leyraud N, Ramshaw B J, Liang R, Bonn D A, Hardy W N and
Taillefer L 2010 \emph{Nature} (London) {\bf 463} 519

\bibitem{Lawler10}
Lawler M J, Fujita K, Lee J, Schmidt A R, Kohsaka Y, Kim C K, Eisaki
H, Uchida S, Davis J C, Sethna J P and Kim E-A 2010 \emph{Nature}
{\bf 466} 347

\bibitem{Chuang10}
Chuang T-M, Allan M-P, Lee J, Xie Y, Ni N, Bud'ko S L, Boebinger G
S, Canfield P C and Davis J C 2010 \emph{Science} {\bf 327} 181


\bibitem{Okazaki11}
Okazaki R, Shibauchi T,  Shi H J, Haga Y, Matsuda T D, Yamamoto E,
Onuki Y, Ikeda H and Matsuda Y 2011 \emph{Science} {\bf 331} 439


\bibitem{Borizi07}
Borzi R A, Grigera S A, Farrell J, Perry R S, Lister S J S, Lee S L,
Tennant D A, Maeno Y and Mackenzie A P 2007 \emph{Science} {\bf 315}
214

\bibitem{Cooper02}
Cooper K B, Lilly M P, Eisenstein J P, Pfeiffer L N and West K W
2002 \emph{Phys. Rev.} B {\bf 65} 241313

\bibitem{Kim08}
Kim E-A, Lawler M J, Oreto P, Sachdev S, Fradkin E and Kivelson S A
2008 \emph{Phys. Rev. B} {\bf 77} 184514

\bibitem{Huh08}
Huh Y and Sachdev S 2008 \emph{Phys. Rev.} B {\bf 78} 064512

\bibitem{Xu08}
Xu C, Qi Y and Sachdev S 2008 \emph{Phys. Rev.} B {\bf 78} 134507

\bibitem{Fritz09}
Fritz L and Sachdev S 2009 \emph{Phys. Rev.} B {\bf 80} 144503

\bibitem{Wang11}
Wang J, Liu G-Z and Kleinert H 2011 \emph{Phys. Rev.} B {\bf 83}
214503

\bibitem{Liu12}
Liu G-Z, Wang J-R and Wang J 2012 \emph{Phys. Rev.} B {\bf 85}
174525

\bibitem{WangLiu}
Wang J and Liu G-Z 2012 arXiv:1205.6164.

\bibitem{Vojta00A}
Vojta M, Zhang Y and Sachdev S 2000 \emph{Phys. Rev.} B {\bf 62}
6721

\bibitem{Vojta00B}
Vojta M, Zhang Y and Sachdev S 2000 \emph{Int. J. Mod. Phys.} B {\bf
14} 3719

\bibitem{Durst}
Durst A C and Lee P A 2000 \emph{Phys. Rev.} B {\bf 62} 1270

\bibitem{Sachdevbook}
Sachdev S 2011 \emph{Quantum Phase Transitions} Chap. 17 (Cambridge
University Press)

\bibitem{heavyfermion}
L\"{o}hneysen H v, Rosch A, Vojta M and W\"{o}lfle P 2007 \emph{Rev.
Mod. Phys.} {\bf 79} 1015

\bibitem{Stockert}
Stockert O, Kirchner S, Steglich F and Si Q 2012 \emph{J. Phys. Soc.
Jpn.} {\bf 81} 011001

\bibitem{Khve}
Khveshchenko D V and Paaske J 2001 \emph{Phys. Rev. Lett.} {\bf 86}
4672

\bibitem{Maris}
Maris P 1995 \emph{Phys. Rev.} D {\bf 52} 6087

\bibitem{Gusynin}
Gusynin V P and Miransky V A 2004 \emph{Eur. Phys.} J. {\bf 37} 363

\bibitem{LiuWang}
Liu G-Z and Wang J-R 2011 \emph{New J. Phys.} {\bf 13} 033022


\end{thebibliography}
\end{document}